\documentclass[%
 reprint,
longbibliography,
 amsmath,amssymb,
 aps,
]{revtex4-1}
\usepackage{lipsum}
\usepackage{graphicx}
\usepackage{dcolumn}
\usepackage{bm}

\usepackage[urlcolor=blue,colorlinks=true,citecolor=blue,breaklinks]{hyperref}
\usepackage[english]{babel}
\usepackage[T1]{fontenc}
\usepackage{amsmath}
\usepackage{hyperref}
\usepackage{amsfonts,amssymb,amsmath}            
\usepackage{lmodern,adjustbox}
\usepackage[skins,breakable]{tcolorbox}
\usepackage{tikz}
\tcbuselibrary{listings}
\tcbuselibrary{breakable}
\usepackage{tikz}
\usepackage{lipsum}
\usepackage{amsmath,amssymb}
\usepackage{esint}
\newcommand{\ket}[1]{|#1\rangle}
\newcommand{\bra}[1]{\langle #1|}
\newcommand{\ketbra}[2]{\ket{#1}\!\bra{#2}}        

\begin{document}

\title{Non-Equilibrium Dynamics of a Dissipative Two-Site Hubbard Model Simulated on IBM Quantum Computers}

\author{Sabine Tornow}
\author{Wolfgang Gehrke}
\author{Udo Helmbrecht}
\affiliation{%
 Research Institute CODE,
Bundeswehr University Munich,
Carl-Wery-Str.~22,
D-81739 Munich}%

\date{\today}

\begin{abstract}
\noindent
Many-body physics is one very well suited field for testing quantum algorithms and for finding wor\-king heuristics on present quantum computers.  We have investigated the non-equilibrium dynamics of one- and two-electron systems, which are coupled to an environment that introduces decoherence and dissipation. In our approach, the electronic system is represented in the framework of a two-site Hubbard model while the environment is modelled by a spin bath. To simulate the non-equilibrium population probabilities of the different states on a quantum computer we have encoded the electronic states and environmental degrees of freedom into qubits and ancilla qubits (bath), respectively.
The total evolution time was divided into short time intervals, during which the system evolves.  After each of these time steps, the system interacts with ancilla qubits representing the bath in thermal equilibrium. We have specifically studied spin baths leading to both, unital and non-unital dynamics of the electronic system and have found that electron correlations clearly enhance the electron transfer rates in the latter case.
For short time periods, the simulation on the quantum computer is found to be in very good agreement with the exact results if error mitigation methods are applied. Our method to simulate also non-unitary time-evolution on a quantum computer can be well extended to simulate electronic systems in correlated spin baths as well as in bosonic and fermionic baths.

\end{abstract}

\maketitle

\section{\label{sec:Int}Introduction}

Recent remarkable technological advances of small quantum computing devices have led to the possibility of testing quantum computing paradigms and finding novel working heuristics. One of the most promising early applications of quantum computers has been the modeling of quantum chemistry systems with many degrees of freedom. 
\cite{McArdle2020, Cao201910856,McCaskey:2019aa,bauer2020quantum}
Available noisy intermediate-scale quantum (NISQ) devices however are still lacking error correction and fault tolerant operations.
\cite{Preskill2018quantumcomputingin}
Nevertheless, exploring in particular the non-equilibrium quantum dynamics of certain open many-body systems has been suggested as a useful near-term application of NISQ devices. The environment may be taken into account by modeling only those of its aspects which would directly affect the system. \cite{Lloyd1073} 
Eventually, the entire environment may be modeled with only one ancilla qubit or even without any ancillas at all, but by utilising the intrinsic noise of the quantum computer itself. \cite{Lloyd1073,rost2020simulation} 

Understanding the dynamics of open many-body systems is important in various fields of fundamental research, including, e.g., unraveling the microscopic mechanism of electron transfer, proton-coupled electron transfer and energy transfer. These key processes in chemistry and biology, with natural energy storage or photosynthesis as prominent examples, feature remarkable energy conversion efficiencies. Advances in understanding such processes directly help engineering important applications such as artificial energy storage and artificial photosynthesis. 
\cite{Marcus,nitzan2013chemical,doi:10.1146/annurev-physchem-040214-121713,kais2014quantum,doi:10.1021/jp208905k}

To simulate the irreversible transfer inherent to these systems any realistic model needs to include vibronic degrees of freedom, e.g., the solvent, the protein or electromagnetic fluctuations leading to 
 dissipation and decoherence.  In this regard,  a paradigmatic model is the spin-boson model \cite{Leggett19871} or its extended versions.
 
 A minimal model, which includes both, electronic correlations and the coupling to an environment introducing dissipation and decoherence, is the two-site dissipative Hubbard model. 
 It has been investigated with the help of real-time path integral Monte Carlo techniques \cite{PhysRevLett.121.110504} and the time-dependent numerical renorma\-lization group. \cite{PhysRevB.78.035434} The main findings were a quantum phase transition, a non-Boltzmann steady state due to a decoherence-free subspace as well as conditions for pair vs. sequential electron transfer.
 
Our aim is exploring possibilities to simulate the dissipative Hubbard model on an available quantum computer.  Although different algorithms for the unitary dynamics of many particle systems modeled by a Hubbard model have been studied and benchmarked before \cite{PhysRevA.92.062318, Barends:2015aa,Las-Heras:2015aa,montanaro2020compressed}, there are only a few reports \cite{PhysRevB.102.125112,Maniscalco,PhysRevB.102.125112,PhysRevResearch.2.023026,Hu:2020aa,dejong2020quantum} for quantum algorithms to si\-mulate the inherent, non-unitary dynamics of dissipative systems. 
In Ref.~\cite{PhysRevA.83.062317}, the authors investigated the dynamics of an open quantum system.  They considered a two-state system without inter-state hopping coupled to a spin bath. Input parameters of the described algorithm were the system environment interaction, the spectral density of the environment as well as the temperature. 
The environment was represented by ancilla qubits designed to have the same effect as the environment in the Hamiltonian.
Ref.~\cite{Hu:2020aa} reports the time evolution of the density matrix, simulated in the operator-sum representation with Kraus operators for the amplitude damping channel. The Kraus operators were converted to a unitary time evolution problem with the help of the Sz.-Nagy dilation theorem.
Ref.~\cite{Maniscalco} discusses a quantum algorithm for many fundamental open quantum systems with unital and non-unital, as well as Markovian and non-Markovian, dynamics. The authors showed that the IBM quantum computer is a robust testbed for implementing a number of theoretical open quantum systems.

Energy transfer has been experimentally simulated using NMR quantum computers. \cite{Wang:2018aa} For this purpose, the photosynthetic system was mapped to the NMR system and the effect of the environment was effectively mi\-micked by a set of pulses, which acted as a classical, pure dephasing noise. Recently, electron and energy transfer have been calculated on a quantum computer. \cite{Guimar_es_2020} In this work, the unitary part of the transfer was calculated on the quantum computer while the dephasing was simulated on a classical computer.

In our work we simulate open many-body systems using IBM quantum computers and have explicitly set out to simulate also the environment on the quantum computer. 
We have considered an open quantum many-body system where the system is described by a two-site Hubbard model. The Hubbard model is mapped onto qubits by a Jordan-Wigner transformation for a filling of one and two electrons.  The time is discretized and a Trotter-Suzuki decomposition is used to calculate the time evolution of the system. The population probabilities are calculated and measured for the total time $t=n \cdot \Delta t$, where $\Delta t$ is the discrete time step and $n$ is the number of gate sequences determining the time evolution of the system.

Our model can be mapped to a coupled two-spin system interacting with a spin bath, which has been calculated on a classical computer for 32 bath spins. \cite{PhysRevE.96.053306} 
In our work, the effective two-spin system is investigated for different couplings to the spin bath which is considered to be in thermal equilibrium, non-interacting and represented by a single ancilla qubit as proposed in Ref.~\cite{PhysRevA.83.062317}.
The XY coupling to the bath state leads to an amplitude damping and non-unital dynamics, whereas the ZZ coupling to the bath state leads to a phase damping and unital dynamics.

We have applied post-processing \cite{Jattana:2020vw,rost2021demonstrating} to mitigate the errors. For short times, the error mitigated results on the quantum computer are found to be in good agreement with the 
exact or classically simulated results. Hence, the IBM quantum computer proves to be a good platform for testing open quantum systems, and - more generally - for performing tests of, and experiments on, novel algorithm implementations and heuristics.

The paper is organized as follows: Sec.~\ref{sec:Model} is devoted to describing our model of the open quantum many-body system and the mapping on qubits and gates. In Sec.~\ref{sec:Qcirq}, we describe the performed experiments on the IBM quantum computer. We summarize our results in Sec.~\ref{sec:Conclusion}  and propose work for future studies.

\begin{figure}[t]

   \includegraphics[width=0.6\linewidth]{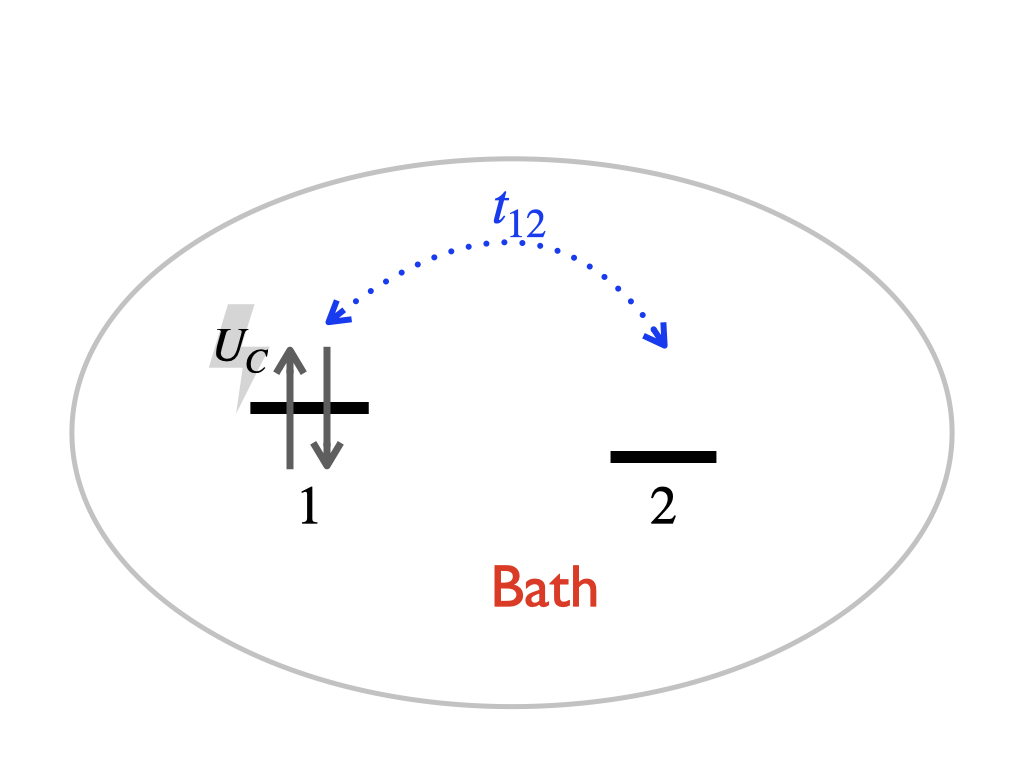}     
      
  \includegraphics[width=0.95\linewidth]{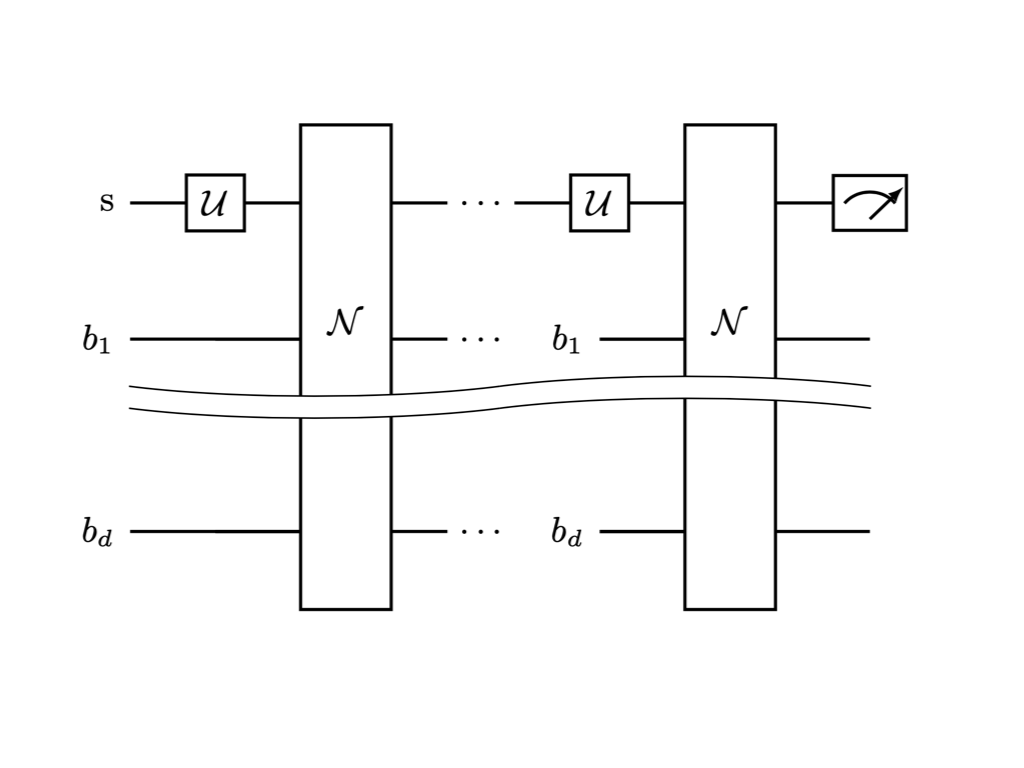}

                \caption{
                Upper panel: Pictorial representation of the Hubbard model with two sites coupled to the bath (environment). Site 1 is occupied with two electrons. The blue dotted arrow indicates the single-electron hopping matrix element $t_{12}$ between the two sites. The on-site repulsion energy of two electrons on the same site is $U_C$. Lower panel: Schematic implementation of trotterized evolution of the open quantum system under study. The system comprises the electronic system qubits $s$ and the bath qubits $b_1$ to $b_d$, which represent the $d$ bath modes of the bath in thermal equilibrium. The time evolution is implemented by $n$ Trotter steps, with the operator $U \approx  (\mathcal{N}\mathcal{U})^n$, where $\mathcal{U}$ implements the time evolution of the system and $\mathcal{N}$ describes the interaction with the bath qubits which were renewed or reset after each step. The reduced state of the electronic system alone is obtained by the indicated measurement. }
                \label{fig:rhotime}
\end{figure}

\section{\label{sec:Model}Model and Methods}

In the following, we will describe our simulation of the dynamics of an open quantum system, which is represented by the Hamiltonian 

\begin{eqnarray}\label{eq:hammodel}
& &H = H_{\textrm{sys}} \otimes \mathbb{I}_{\textrm{bath}}+ \mathbb{I}_{\textrm{sys}} \otimes H_{\textrm{bath}}+ H_{\textrm{sys-bath}}.
\end{eqnarray}
where $H_{\textrm{sys}}$, $H_{\textrm{bath}}$ and $H_{\textrm{sys-bath}}$ are, respectively, the system, bath and interaction Hamiltonians, and $\mathbb{I}$ is the identity operator.

While it is more common to consider a bath of harmonic oscillators, \cite{PhysRevB.78.035434} we choose here instead a spin bath for the quantum computer. This spin bath is fully characterized by the spectral function $J(\omega)$, depending on the distribution of frequencies $\omega$ and couplings. The influence of the system on the environment is neglected.

It was shown in Ref.~\cite{PhysRevA.83.062317} that a large environment can be simulated by resetting the bath into its thermal equilibrium state during the calculation. The spectral density $J(\omega)$ is discretized in different modes, each of which is represented by a single qubit. 

We consider the two-site Hubbard model filled with one or two electrons coupled to an environment or bath, as depicted in Fig.~\ref{fig:rhotime} (upper panel). We assume that the system and bath are initially in a pro\-duct state of density matrices $\rho_{\rm sys} \otimes \rho_{\rm bath}$. Since we are only interested in the dynamics of the electronic system, we partially trace out the bath degrees of freedom leading to the completely positive map 
\begin{eqnarray}
	\rho \rightarrow {\rm Tr}_{\rm bath} \left( U (\rho_{\rm sys} \otimes \rho_{\rm bath}) U^{\dag}\right)
\end{eqnarray}

which describes the density matrix of the electronic system at $t$, and $U=e^{-i H t}$ is the unitary time evolution operator imposed on the total electron-bath system.

To set up the full model, we need one or two system qubits $s$ and bath qubits $b_1$ to $b_d$ (see Fig.~\ref{fig:rhotime} (lower panel). In general, already a single ancilla qubit could be sufficient to represent the entire environment.
The time evolution in the Trotter form is implemented by the operator $U(\Delta t)=\mathcal{N}(\Delta t)\mathcal{U}(\Delta t)$, where $\mathcal{U}(\Delta t)$ implements the time evolution of the system and $\mathcal{N}(\Delta t)$ describes the interaction with the bath qubits, representing the bath in thermal equilibrium. The density matrix $\rho_{\rm bath}$ is forced back into the thermal equilibrium with a reset of the bath qubits. Alternatively, fresh qubits can be chosen. While the first method leads to an amplitude damping error during the time of the reset operation \cite{rost2021demonstrating} the second method leads to a fidelity drop since more distant fresh bath qubits require a large number of swapping operations. Both errors can be mitigated with an appropriate method for short quantum circuits. \cite{Jattana:2020vw,rost2021demonstrating}
This procedure is repeated $n$ times and the occupation probability is measured during the time $t=n \cdot \Delta t$.  After each interaction with the bath its degrees of freedom are traced out.  
This is very similar to the
 collision approach where decoherence and dissipation are introduced by repeated collisions between system and bath qubits during a time interval $\Delta t$. \cite{Maniscalco,cattaneo2020collision}
 In the following chapter we first discuss the electronic system. 
 
\subsection{\label{subsec:ElSys}Electronic System}

In the electronic system we consider a two-site Hubbard model (Fig~\ref{fig:rhotime} (upper panel))
\begin{eqnarray}\label{eq:hamsys}
& &H_{\textrm{sys}} = \sum_{\sigma,i=1,2} \varepsilon_i~c_{i,\sigma}^{\dag}c_{i,\sigma} +t_{12} \sum_{\sigma}\left(c_{2,\sigma}^{\dag}c_{1,\sigma}+c_{1,\sigma}^{\dag}c_{2,\sigma} \right) \nonumber\\
& &+ U_C \sum_{i=1,2} c_{i\uparrow}^{\dagger}c_{i\uparrow}c_{i\downarrow}^{\dagger}c_{i\downarrow}
\end{eqnarray}

where $ c_i \ (c_i^{\dagger})$ denotes a fermionic annihilation (creation) operator for electrons on site $i=1,2$ with spin $\sigma$.  $\epsilon_i$ is the onsite energy, $t_{12}$ depicts the hopping matrix element between the two sites and $U_C$ is the onsite Coulomb repulsion.  

The Hilbert space can be divided into different subspaces. With a filling of one electron we effectively deal with a two-state system with $\ket{0}=\ket{\uparrow,0}$ and $\ket{1}=\ket{0,\uparrow}$.
The subsystem with two electrons and its total spin equal to zero is spanned by the states $\ket{00}=\ket{\uparrow \downarrow,0}$, 
$\ket{01}=\ket{\downarrow,\uparrow}$, $\ket{10}=\ket{\uparrow,\downarrow}$ and $\ket{11}=\ket{0,\uparrow \downarrow}$ with the notation $\ket{\rm site 1, site2}$ describing the occupation on site $1$ and site $2$, respectively.
We discuss both subspaces in more detail in the following.

\subsubsection{Filling with one electron}

In the subspace with one electron, which is spanned by $\ket{0}$ and  $\ket{1}$, we can rewrite the Hamiltonian 
\begin{eqnarray}
H_{\textrm{sys}}=\varepsilon ~ \sigma_z+ t_{12} ~ \sigma_x,
\label{eq:Hsys}
\end{eqnarray}
with $\varepsilon_1=-\varepsilon$, $\varepsilon_2=\varepsilon$,
and $\sigma_x$, $\sigma_z$ are the Pauli matrices.

To simulate the time evolution of the population probabilities
we discretize the time $t= n \cdot \Delta t$, write the time evolution operator as 
\begin{eqnarray}
e^{-iHt}=\left(e^{-iH \Delta t}\right)^n	
\end{eqnarray}

and trotterize the Hamiltonian of the electronic system into the non-commuting parts $\varepsilon \cdot \sigma_z$ and $t_{12} ~\cdot \sigma_x$

\begin{eqnarray}
U(t)=e^{-i H_{\rm sys} t}\approx  \left(e^{-i t_{12} \cdot \sigma_x \cdot \Delta t}e^{-i \epsilon \cdot \sigma_z \cdot \Delta t}\right)^n \ .
\end{eqnarray}

We can approximate the time evolution operator for small $\Delta t$ :
\begin{eqnarray}
	U(n \cdot \Delta t)\approx \left(R_x(2\cdot t_{12} \cdot \Delta t) \cdot R_z(2\cdot \epsilon \cdot \Delta t) \right)^n
\end{eqnarray}
with
\begin{eqnarray}	
R_x(2\cdot t_{12}\cdot \Delta t)  &=& e^{ -i t_{12} \cdot \sigma_x \cdot \Delta t} \\
&=&
    \begin{pmatrix}
        \cos(t_{12} \cdot \Delta t)   & -i \sin(t_{12} \cdot \Delta t) \\
        -i \sin(t_{12} \cdot \Delta t) & \cos(t_{12} \cdot \Delta t)
    \end{pmatrix} \nonumber
 \end{eqnarray}
    and
\begin{eqnarray}	
R_z(2\cdot \epsilon \cdot \Delta t) 
&=& e^{-i \epsilon \cdot \sigma_z \cdot \Delta t} \\
&=&    \begin{pmatrix}
        e^{-i \epsilon \cdot \Delta t} & 0 \\
        0 & e^{i \epsilon \cdot \Delta t}
    \end{pmatrix} \nonumber
  \end{eqnarray}
  
and implement it on the quantum computer.

\subsubsection{Filling with two electrons}

In the subspace with two electrons which is spanned by $\ket{0}$, $\ket{1}$, $\ket{2}$ and $\ket{3}$ we can rewrite the Hamiltonian in the Pauli matrix form
\begin{eqnarray}
H=H_{t}+H_U
\end{eqnarray}
with 
\begin{eqnarray}
H_{t}=t_{12} \ \left(\mathbb{I}_2 \otimes \sigma_x + \sigma_x \otimes \mathbb{I}_2 \right)
 \end{eqnarray}
 and 
 \begin{eqnarray}
H_U=\frac{U_c}{2} \left(\sigma_z \otimes \sigma_z \right)
\end{eqnarray} 

\noindent
We restrict ourselves to $\epsilon=0$.  
The time evolution operator is approximately decomposed (Trotter-Suzuki) in the product form
\begin{eqnarray}
U(t) \approx \left(e^{-i H_{t} \Delta t} \cdot  e^{-i H_{C} \Delta t}\right)^n
\label{eq:TrotterTwo}
\end{eqnarray}
with 
 \begin{eqnarray}
e^{-i H_{t} \Delta t} &=& e^{-i t_{12} \cdot \sigma_x \Delta t}   \otimes
e^{-i t_{12}\cdot \sigma_x \Delta t} \\
&=&R_x(2 \cdot t_{12} \cdot \Delta t) \otimes R_x(2\cdot t_{12} \cdot \Delta t)
\end{eqnarray}
The second factor of Eq.~(\ref{eq:TrotterTwo}) can be written as:
\begin{eqnarray}
\label{eq:twoel}
R_{ZZ}(U_C \cdot \Delta t)&=&e^{-i H_{C} \Delta t}=e^{-i \frac{U_c}{2} (\sigma_z \otimes \sigma_z) \Delta t} 
\\ &=&
    \begin{pmatrix}
        e^{-i\frac{U_C \Delta t}{2}} & 0 & 0 & 0 \\
        0 & e^{i\frac{U_C \Delta t}{2}} & 0 & 0 \\
        0 & 0 & e^{i\frac{U_C \Delta t}{2}} & 0 \\
        0 & 0 & 0 & e^{-i\frac{U_C \Delta t}{2}}
    \end{pmatrix} \nonumber
    \end{eqnarray}
which is $R_{ZZ}(U_C \cdot \Delta t)=U_{\rm CNOT} (\mathbb{I}_2 \otimes R_Z(U_c \cdot \Delta t)) U_{\rm CNOT}$, leading to the circuit representation of the unitary time evolution operator 
\begin{eqnarray}
	U(n \cdot \Delta t)\approx \left(R_{ZZ}(U_C \cdot \Delta t) (R_x(2\cdot t_{12} \cdot \Delta t))^{\otimes 2}  \right)^n
\end{eqnarray}
In the next section we discuss the bath and the coupling to the bath.

\subsection{\label{subsec:Env}Environment}

For the environment we consider a bath of independent two-level systems (spins) in thermal equilibrium. 
We consider two different models for the bath and its coupling to the system, first, for the ZZ coupling
\begin{eqnarray}
	H_{{\rm sys-bath}}^{ZZ}&=&\sum_{k=1}^{d}g_k \ (\sigma_z \otimes \sigma_{z,k} ) \\
	H_{\rm bath}^{ZZ}&=&-\sum_{k=1}^d\omega_k \sigma_{x,k} 
	\label{eq:bathZ}
\end{eqnarray}
and second, for the XY coupling
\begin{eqnarray}
	H_{{\rm sys-bath}}^{XY}&=&\sum_{k=1}^{d}  g_k \ (\sigma_{x} \otimes \sigma_{x,k} +\sigma_{y} \otimes \sigma_{y,k}) \\
	H_{\rm bath}^{XY}&=&-\sum_{k=1}^{d}\omega_k \sigma_{z,k}
	\label{eq:bathXY}
\end{eqnarray}
\noindent where $\sigma_{z}$ $(\sigma_{z,k})$ is the system spin (bath spin), $\omega_k$ are the frequencies of the environment mode, $g_k$ is the coupling coefficient for the ${\rm k^{th}}$ mode and $d$ is the number of bath modes.

We describe the electron dynamics by the reduced density matrix
and assume that the initial density matrix is a tensor product of the system and the bath density matrices. The bath degrees of freedom are in their thermal equilibrium. 
We discretize the time in $t= n \cdot \Delta t$ and at every time step we apply the unitary time evolution operator 

\begin{eqnarray}
	U(\Delta t)= \mathcal{N} (\Delta t) \mathcal{U} (\Delta t)
\end{eqnarray}
During the time interval $\Delta t$ the electronic system interacts locally with the degrees of freedom of the environment, i.e., the ancilla qubits as shown in Fig.~\ref{fig:rhotime} (lower panel). For the population probability at time $t$ we need $n \cdot d$ ancilla qubits. Here we use the approximation that the whole bath is represented by one qubit, so that we have to reset the one (two) bath qubit(s) $n$ times or use $n$ $(2n)$ fresh qubits in the single (two) electron case, respectively.
We are only interested in the degrees of freedom of the electronic system and calculate the reduced density operator: 
\begin{eqnarray}
\rho(n) \rightarrow 
{\rm Tr}_{\rm bath} \left( (\mathcal{N }\mathcal{U} )^n    (\rho_{\rm System} \otimes  \rho_{\rm bath} ) (
\mathcal{U}^{\dagger}\mathcal{N}^{\dagger} )^n \right) \nonumber 
\end{eqnarray}

The electronic system evolves for the time $\Delta t$ with the unitary time evolution

\begin{eqnarray}
	\mathcal{U}=e^{-i ~H_{sys} ~\Delta t}
\end{eqnarray}

and interacts with one of the ancilla qubits which represents the entire bath.
The unitary time evolution for the ZZ coupling is
\begin{eqnarray}
	\mathcal{N}^{ZZ}=e^{-i  ~g ~ (\sigma_z \otimes \sigma_{z}) ~\Delta t}
\end{eqnarray}
The bath qubit is in thermal equilibrium which can be emulated in this case by putting the qubits into the state $\rho_{bath}=\ketbra{+}{+}$.

The operator $\mathcal{N}^{ZZ}$ acts as follows:
\begin{eqnarray}
	\mathcal{N}^{ZZ} (\ket{0}_{\rm sys} \otimes \ket{0}_{\rm bath})&=&e^{\frac{i g \Delta t}{2}}(\ket{0}_{\rm sys} \otimes \ket{0}_{\rm bath}) \\
	\mathcal{N}^{ZZ}  (\ket{0}_{\rm sys} \otimes \ket{1}_{\rm bath})&=&e^{-\frac{i g \Delta t}{2}}(\ket{0}_{\rm sys} \otimes \ket{1}_{\rm bath}) \\
	\mathcal{N}^{ZZ}  (\ket{1}_{\rm sys} \otimes \ket{0}_{\rm bath})&=&e^{-\frac{i g \Delta t}{2}}(\ket{1}_{\rm sys} \otimes \ket{0}_{\rm bath}) \\
	\mathcal{N}^{ZZ}  (\ket{1}_{\rm sys} \otimes \ket{1}_{\rm bath})&=&e^{\frac{i g \Delta t}{2}}(\ket{1}_{\rm sys} \otimes \ket{1}_{\rm bath}) 
\end{eqnarray}

For the second case (the $XY$ interaction) the time evolution is given by 
\begin{eqnarray}
	\mathcal{N}^{XY}=e^{-i  g  (\sigma_x \otimes \sigma_x+\sigma_y \otimes \sigma_y) \Delta t} \ .
\end{eqnarray}
The operator $\mathcal{N}^{XY}$ acts as follows: 
\begin{eqnarray}
	\mathcal{N}^{XY} (\ket{0}_{\rm sys} \otimes \ket{0}_{\rm bath})&=&\ket{0}_{\rm sys} \otimes \ket{0}_{\rm bath} \\
	\mathcal{N}^{XY} (\ket{1}_{\rm sys} \otimes \ket{0}_{\rm bath})&=&\sin(g \Delta t)(\ket{0}_{\rm sys} \otimes \ket{1}_{\rm bath}) \nonumber \\
	&+& \cos(g \Delta t)(\ket{1}_{\rm sys} \otimes \ket{0}_{\rm bath}) \nonumber
\end{eqnarray}

After each collision, the electronic system interacts with the next ancilla qubit in the state $\rho_{bath}=\ketbra{0}{0}$.

\subsection{\label{subsec:ErrorMitigation}Error Mitigation}
\subsubsection{General error mitigation method}
Errors are due to
noise, decoherence and dissipation as well as they occur during read-out. They can be coherent, incoherent or correlated.
To improve the experimental results we adapt the general error mitigation method proposed in Ref. \cite{Jattana:2020vw}, which was developed to get a reasonable result recovery independent of the circuit depth, gate sets and error type. 
In the following we apply the error mitigation methodology only to the system qubits. 
We prepare calibration circuits for the initial states $\ket{0}$, $\ket{1}$ ($\ket{00}$, $\ket{01}$, $\ket{10}$, $\ket{11}$)
for each discrete time step $n$ for the one (two) electron case, respectively. We consider for each calibration circuit at Trotter step $n$ the actual circuit at Trotter step  $n/2 $ (for $n$ even) or $(n-1)/2$ (for $n$ odd) and add the inverse, so that the unitary of the circuit is equivalent to the identity matrix.  

The calibration matrix $M_Q$ for each Trotter step $n$ consists of the measurement results of the $2$ ($4$) calibration circuits in the one (two) electron case. The mitigated results are obtained by minimizing the function
\begin{eqnarray}
    f(\vec{x})=\sum_{i=1}^{N} \left(v_i-(M_Q \ \vec{x})_i \right)^2
\end{eqnarray}
with the constraint $0 \leq x_i \leq 1$ and $\sum_i x_i=1$ for all $x_i$. Here, $\vec{v}$ are the experimental raw data and $\vec{x}$ are the error mitigated results.

\subsubsection{Mitigation of the amplitude damping error during qubit reset}
In addition, errors occur during the reset operation.
The reset is performed by measuring the state of the qubits and applying a conditional $X$ gate. The duration of the reset operation is about an order of magnitude longer than the duration of the other operations during one Trotter step, eventually leading to a relaxation $\ket{1} \rightarrow \ket{0}$ and therefore to amplitude damping. To mitigate the noise,
 we perform the experiment for $2, 3$ and $4$ resets instead of one at each Trotter step and extrapolate to the zero reset-time limit which was proposed in Ref. \cite{rost2021demonstrating}.

%

\section{\label{sec:Qcirq} Quantum Simulation of the non-equilibrium dynamics }
We simulated the dissipative Hubbard model 
on the 27 qubits IBM Q devices {\it Montreal}, {\it Toronto} and {\it Paris}. We performed five experiments with 8192 runs for each model circuit as well as for the 2 (4) calibration circuits for the single (two) electron case. 
For the bath qubit, we used either fresh ancilla qubits in the default initial state $\ket{0}$ or we used the reset option, which is performed $1$, $2$, $3$, $4$, $5$ times to perform a zero reset-time extrapolation. To trace out the bath, we measure the one or two system qubits at the end of the circuit sequence. 
 
The measurement results of the experimental quantum circuits implementations on the quantum computer are compared for every Trotter step $n$ to the error-mitigated results and to the results of the classical simulations without noise as well as to the classical simulations of the Lindblad master equation for continuous times.

 \subsection{Filling with one electron}

 We simulate the two state system in the amplitude and phase damping channel on the IBM quantum computer. The time-evolution of the occupation probabi\-lities is obtained by measuring the expectation value $n_1={\rm Tr} \left(\ketbra{0}{0}  \rho(t)\right)$ and 
$n_2={\rm Tr} \left(\ketbra{1}{1}  \rho(t)\right)$ with the initial state $\rho(0)=\ketbra{1}{1}$ for every Trotter step $n$. 
We benchmark the experimental results with the numerical calculation of the discrete Trotter evolution as well as the continuous time simulation of the Lindblad master equation 
 \begin{eqnarray}
 \label{lindbladone}
\dot\rho(t)&=&-i[H_{\rm sys},\rho(t)]\\ \nonumber
& +&\gamma   \left[ C \rho(t) C^{\dag} -\frac{1}{2} \{C^{\dag} C,\rho(t) \} \right]
\end{eqnarray}
 with $H_{\rm sys}=\varepsilon ~ \sigma_z+ t_{12} ~ \sigma_x$ 
 where we chose $C=\sigma^{+}$ ($C^{\dag}=\sigma^{-}$) and $\sigma^{\pm}=\sigma_x\pm i \sigma_x $) for the amplitude damping and $C=\sigma_z$ for the phase damping channel, respectively. ( $[\cdot, \cdot]$ is a commutator and the $\{\cdot, \cdot \}$ is an anticommutator.)

We chose the Trotter step size $\Delta t=\frac{t}{n}=0.8$ to minimize the Trotter error until $n=25$ in comparison to the continuous time calculation. The other parameters are $ t_{12}\cdot \Delta t=0.2$ as well as $\epsilon \cdot \Delta t=0.8$ ($\epsilon \cdot \Delta t=0$) for the finite (zero) bias case.

 \begin{widetext}
\onecolumngrid
 \begin{figure}[t]
             \includegraphics[width=0.9\linewidth]{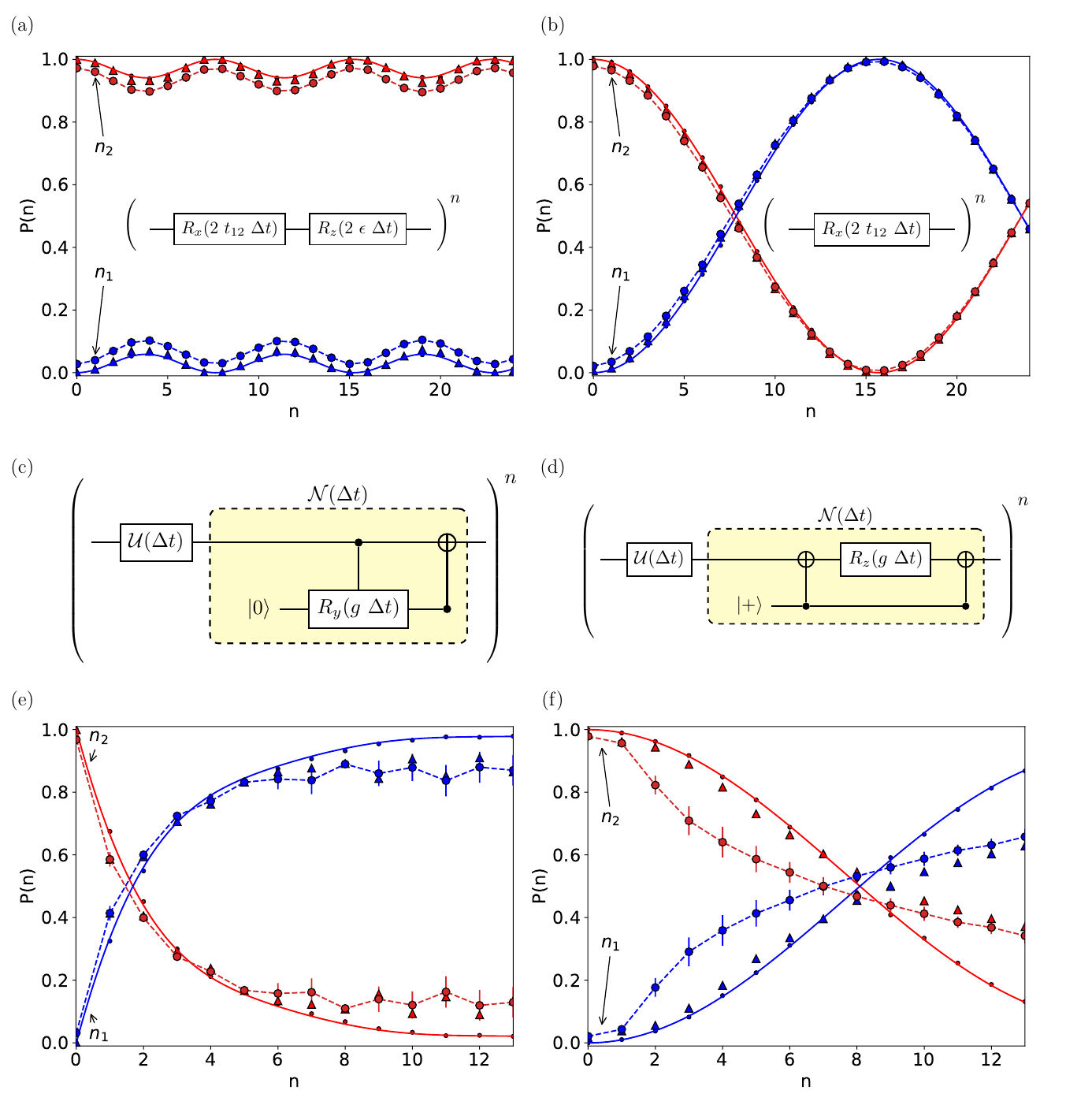}
                \caption{Circuit implementation and results of the free and damped one-electron time evolution. The graphs show the population probabilities $P(n)=n_1$ (blue) and $n_2$ (red) as a function of the number of discrete time steps $n$: classical simulation of the Lindblad master equation (solid line), classical simulation of the quantum circuit (dots), experiment on the QC (dashed line, circles) and error mitigated experiment on the QC (triangles). The hopping parameter is $t_{12} \cdot \Delta t=0.2$. If not otherwise stated we apply the general error mitigation. The gate sequence is repeated $n$ times for the time $t=n \cdot \Delta t$.
                (a) Population probabilities with $\epsilon \cdot \Delta t=0.8$. Inset:  
                Circuit implementation of the single-electron time evolution with discrete evolution operators $R_x(2~t_{12}) \cdot \Delta t) R_z(2~\epsilon \cdot \Delta t)$. 
                (b) Population probabilities with $\epsilon \cdot \Delta t=0.0 $. Inset:  
                Circuit implementation of the single-electron time evolution with discrete evolution operators $R_x(2~t_{12}$. 
                (c) Circuit implementation for the amplitude damping channel (XY coupling). (d) Circuit for the phase damping channel (ZZ coupling). (e) Results of (c).
                 The parameters are $\epsilon \cdot \Delta t=0.8 $, $g \cdot \Delta t= 1.28$ and $\gamma=1.0$ (XY coupling). 
                (f) Results of (d). The parameters are $\epsilon \cdot \Delta t=0 $, 
                 	$ g \cdot \Delta t= 0.32$ and $\gamma=0.19$ (ZZ coupling). As error mitigation, in this case, the zero reset-time extrapolation is applied.
    }
                \label{fig:singleEl}
\end{figure}
\end{widetext}
\twocolumngrid 

Before simulating the non-unitary time evolution of the two state system in the spin bath we display the unitary time evolution of the system alone in
 Fig.~\ref{fig:singleEl} (a) and (b) where the well-known Rabi oscillations can be seen. The error mitigated experimental result on the quantum computer coincides with the theoretical curve, which can be obtained analytically from eq. (\ref{lindbladone}) for $\gamma=0$ as well as with the classical Trotter simulation of the circuit. The trotterized quantum circuit with an $R_z$ gate (Fig.~\ref{fig:singleEl} (a)) does not introduce additional noise since it is a virtual gate. \cite{McKay2017}
 
The circuit implementation of the time evolution operator $U(\Delta t) =  \mathcal{N} (\Delta t)\mathcal{U} (\Delta t)$, where $\mathcal{U} (\Delta t)$ implements the time evolution of the electronic system in Fig.~\ref{fig:singleEl} (a) and $\mathcal{N} (\Delta t) $ describes the interaction with the bath qubits, is displayed in Fig.~\ref{fig:singleEl} (c). Fig.~\ref{fig:singleEl} (d) displays the circuit for the phase damping channel with $ZZ$ coupling.

In Fig.~\ref{fig:singleEl} (e) we introduce the amplitude damping channel. 
After multiple interactions, the process leads to amplitude damping and $\ket{1} \rightarrow \ket{0}$.
Counteracting to this decay is the hopping between the sites. 
Since the effective hopping $t_{12}^2/\epsilon$  for $\epsilon \gg t_{12}$ is small the transfer is faster in comparison to the case $\epsilon < t_{12}$. 
Both the qubit reset and use of fresh qubits give a similar result for small time steps until $n=8$. 
 
 The phase damping circuit Fig.~\ref{fig:singleEl} (d) leads to decoherence and damped Rabi oscillations, see Fig.~\ref{fig:singleEl} (f). 
 The electronic system evolves with $\mathcal{U} (\Delta t)$. Then the system interacts with an ancilla qubit
 in the initial state $\ket{+}$. The electronic system is measured at $t= n \cdot \Delta t$ after $n$ collisions ($n$ repetitions of the circuit). The successive collisions introduce a phase damping $\rho \rightarrow \frac{1}{2} \mathbb{I}_2$ for $t \rightarrow \infty$.
 We chose a small coupling $g \cdot \Delta t = 0.16$ and reset the bath qubit at each Trotter step. The error mitigated curve lies close to the theoretical curve until $n=8$. For larger values of $n$ the errors are not mitigated anymore. Besides dissipation and decoherence a source of the error is the decay of $|1 \rangle$ into the state $|0 \rangle$ during the reset. When using (instead the reset of a qubit) fresh qubits, multiple SWAP gates have to be introduced leading to a strong decoherence. Therefore, the population probabilities equilibrate to $0.5$ much earlier and therefore the ancilla qubit reset method with the zero reset-time extrapolation error mitigation is favorable.

The computation shows that it is in principle possible to simulate open quantum one-electron systems on the quantum computer with error mitigation, for short time scales.

\begin{figure}[t]
            \hspace{-0cm}
                \includegraphics[width=1\linewidth]{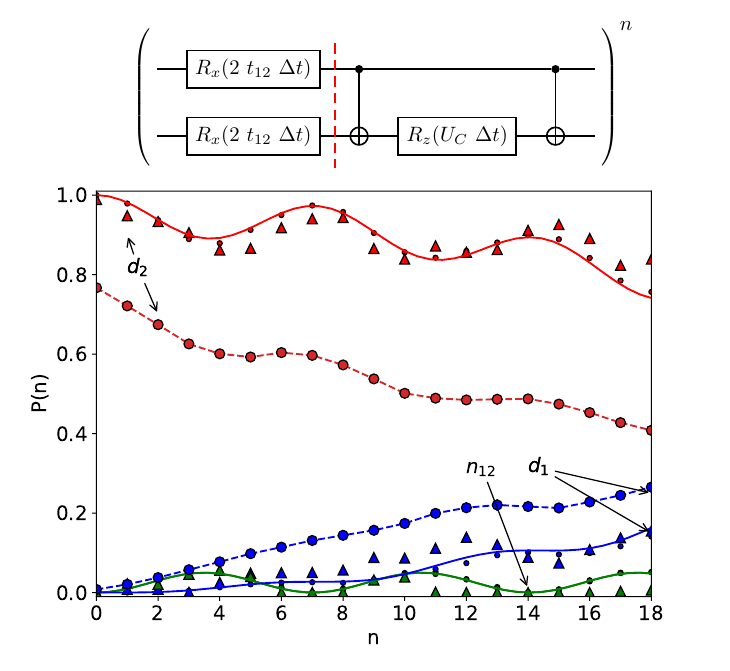}
                \caption{Circuit implementation (upper panel) and results of the two-electron time evolution. Lower panel:  population probabilities $P(n)=d_1$ (blue), $d_2$, (red) and $n_{12}$ (green) as a function of the number of discrete time steps $n$, exact (solid line); classical simulation of the above quantum circuit (dots), experiment on the QC (dashed line, circles) and error mitigated experiment on the QC (triangles).
                	The parameters are $t_{12} \cdot \Delta t=0.2$, $U_C \cdot \Delta t=0.8$ and 
                	$g \cdot \Delta t= 0.0$.
                	The general error mitigation is applied.
                }
                \label{fig:twoEl}
\end{figure}

\begin{figure}[t]
            \hspace{-0cm}

         \includegraphics[width=1
                \linewidth]{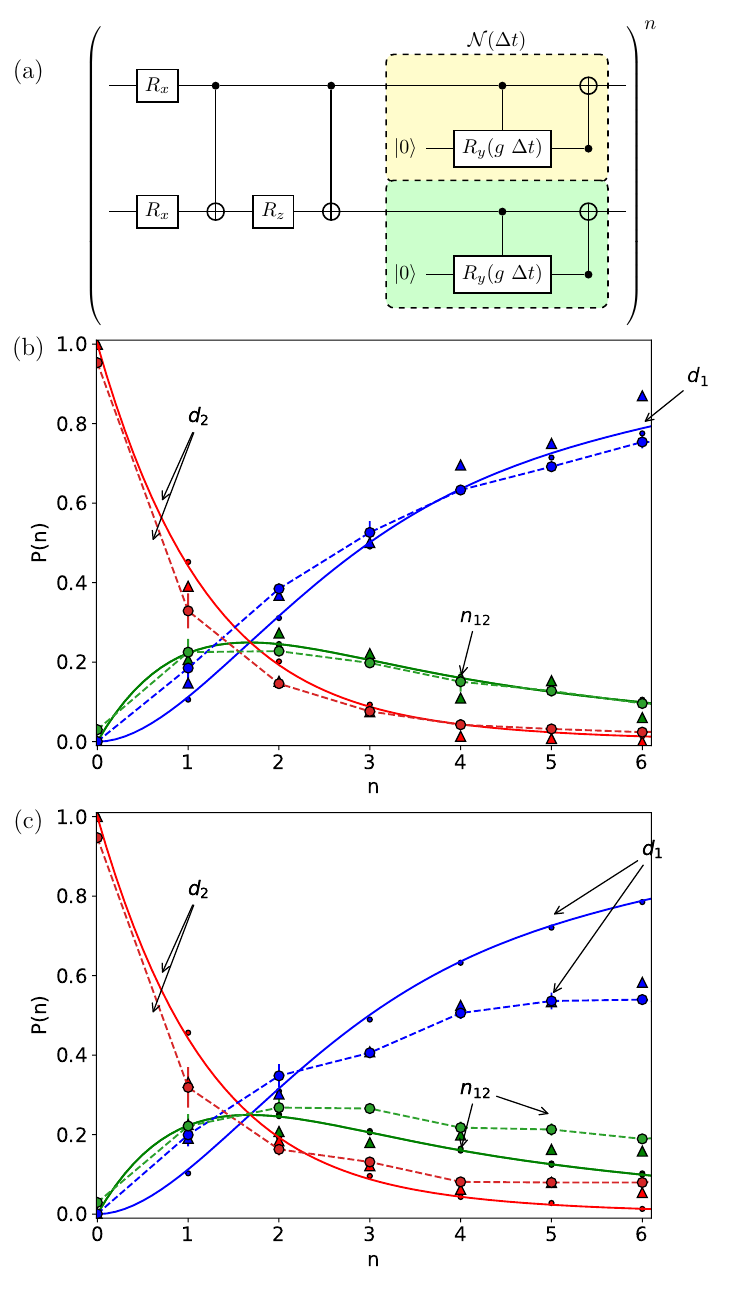}
                \caption{Circuit implementation and results of the damped two-electron time evolution. (a) Circuit implementation for the amplitude damping channel (XY coupling). (b) and (c): Population probabilities $P(n)=d_1$ (blue), $d_2$, (red) and $n_{12}$ (green) as a function of the number of discrete time steps $n$; classical simulation of the Lindblad master equation with $\gamma=1.0$ (solid line), classical simulation of the quantum circuit (dots), experiment on the QC (dashed line, circles) and error mitigated experiment on the QC (triangles).
                	The parameters are $t_{12} \cdot \Delta t=0.2$, $U_C \cdot \Delta t=0.8$ and 
                	$ g  \cdot \Delta t= 0.64$ (XY coupling).
                	The bath in thermal equilibrium is modeled by a reset of the bath qubits (b) and fresh qubits (c).}
               \label{fig:twoEldamped}
\end{figure}

\subsection{Filling with two electrons}

 We simulate the four state system in the amplitude damping channel on the IBM quantum computer. The time-evolution of the occupation probabi\-lities is obtained by measuring the expectation values $d_1={\rm Tr} \left(\ketbra{00}{00}  \rho(t)\right)$, 
 $d_2={\rm Tr} \left(\ketbra{11}{11}  \rho(t)\right)$,  $n_{12}={\rm Tr} \left(\ketbra{01}{01}  \rho(t)\right)$ and $n_{21}={\rm Tr} \left(\ketbra{10}{10}  \rho(t)\right)$
 with the initial state $\rho(0)=\ketbra{11}{11}$ for every Trotter step $n$ of the quantum circuits.
We benchmark the experimental results with the numerical calculation of the discrete Trotter evolution as well as with the continuous time simulation of the Lindblad master equation

\begin{eqnarray}
\label{lindbladtwo}
\dot\rho(t)&=&-i[H_{\rm sys},\rho(t)]+\\ \nonumber
& +&\gamma  \sum_{n=1}^2 \left[ C_n \rho(t) C_n^{\dag} -\frac{1}{2} \{C_n^{\dag} C_n,\rho(t) \} \right]
\end{eqnarray}

with $C_1= \mathbb{I}\otimes \sigma^{+}$ and $C_2=\sigma^{+} \otimes \mathbb{I}$ and     
\begin{eqnarray}
H_{sys}=t_{12} \ \left(\mathbb{I}_2 \otimes \sigma_x + \sigma_x \otimes \mathbb{I}_2 \right)
+\frac{U_c}{2} \left(\sigma_z \otimes \sigma_z \right)
\end{eqnarray}

Before simulating the two state system in the spin bath we calculate the system dynamics alone, see Fig. \ref{fig:twoEl}. 
The circuit depicted in Fig. \ref{fig:twoEl} (upper panel) is split into the kinetic part $R_x(2~t_{12} \cdot \Delta t) \otimes R_x(2~t_{12} \cdot \Delta t) $ before the dashed line and the Coulomb interaction part $R_{ZZ}(U_C \cdot \Delta t)$, see
	eq.~(\ref{eq:twoel}), after the dashed line. The gate sequence is repeated $n$ times for the time $t=n \cdot \Delta t$.
The two-electron dynamics (Fig. \ref{fig:twoEl} (lower panel)) shows pair oscillations: The electron pair oscillates between site 1 and 2 with the small frequency $4 t_{12}^2/U_C$ whereas the fast oscillations with frequency $U_C$ characterize the virtual population of the low lying states $\ket{10}$ and $\ket{01}$.  \cite{PhysRevB.78.035434} The error mitigated experimental result on the quantum computer coincides with the theoretical curve, which can be obtained analytically from eq. (\ref{lindbladtwo}) for $\gamma=0$. Although the experimental raw data show a large deviation from the theoretical curve, the general error mitigation scheme turns out to be very effective in recovering the result without noise.

We have finally investigated the effect of dissipation.
The corresponding quantum circuit with the amplitude damping channel (\ref{fig:twoEldamped} (a)) of the time evolution operator $U(\Delta t) = \mathcal{N} (\Delta t)\mathcal{U} (\Delta t)$ now comprises not only the time evolution of the electronic system $\mathcal{U} (\Delta t)$ (Fig. \ref{fig:twoEl} (upper panel)) but also the interaction $\mathcal{N} (\Delta t) $ with the bath qubits. 
The amplitude damping channel is tested with the qubit reset method and the fresh qubit method (cf. \ref{fig:twoEldamped} (b) and (c), respectively). Irrespective of the employed method, the initial state ($\ket{11}$) is decaying very fast towards the thermal equilibrium due to the amplitude damping and small effective pair hopping $t_{12}^2/U_C$. Eventually, the two electrons are transferred to the other site irreversibly.
Comparing both methods however, it is clearly seen that the qubit reset method is more accurate than the fresh qubit method, especially for $n>2$.

We demonstrate that open quantum systems like photosystem models can be well simulated on a quantum computer. While we have used a minimal model including both electron correlation and a coupling to a bath,
our method can be extended to more realistic models with more states as well as to different also non-Markovian baths to model realistic photosystems if error mitigation methods are used. 
\section{\label{sec:Conclusion}Conclusion}

We have studied quantum algorithm heuristics for si\-mulating the time evolution of an open quantum system with an environment, which introduces decoherence and dissipation, on an IBM quantum computer.  We simulated a two-site dissipative Hubbard model in the single and two-electron subspace, which were mapped onto one and two system qubits, respectively. The entire spin bath in thermal equilibrium is represented by one or two ancilla qubits \cite{PhysRevA.83.062317}. This model is equivalent to an interacting two-spin system coupled to a non-interacting spin-bath. 
As proposed in Refs. \cite{cleve2019efficient,Maniscalco,cattaneo2020collision}, we have implemented our system-bath interaction as "consecutive collisions" between system and bath qubits. The system evolves for the small time interval $ \Delta t$ during which it interacts with the bath in thermal equilibrium, with the strength $g$. After each time step $n$ we trace out and disregard the bath degrees of freedom. 

With the coupling to the bath decoherence and dissipation are introduced. Two different couplings to bath qubits were considered: The ZZ coupling to the spin-bath qubit leads to a phase damping which in turn leads to decoherence and unital dynamics. This evokes damping of the Rabi oscillations. 
The XY coupling to the spin-bath qubit leads to dissipation and non-unital dynamics so that the system will irreversibly end up in state $\ket{0}$ or $\ket{00}$, respectively. Counteracting is the hopping matrix element which forces the system to go back to the other states. The effective hopping gets small for a large Coulomb interaction so that the dissipative environment in addition to the electronic correlations leads to a fast and irreversible electron-pair transfer. 

 The experimental data for the one- and two-electron system are found to be in very good agreement with the theoretical or analytical result for small $n$, if the ge\-neral error mitigation method of Ref. \cite{Jattana:2020vw} and the error mitigation with the zero reset-time extrapolating \cite{rost2021demonstrating} are applied.

In summary, we have demonstrated that today's emer\-ging quantum computing hardware enables well the eva\-luation of quantum algorithms for open many-particle systems. In future studies, we will investigate other important environment models and couplings, e.g., ancilla qubits which simulate a bath of harmonic oscillators, a fermionic bath or correlated bath qubits to simulate non-Markovian processes. 

\hspace{1cm}

\section*{Acknowledgments}
The authors thank D.~J.~Egger, L.~Del Re, B.~Rost, A.~Kemper and J.~Freericks for insightful discussions.
The results presented in this paper were obtained in part using an IBM Q quantum computing system as part of the IBM Q Network. The views expressed are those of the authors and do not reflect the official policy or position of IBM or the IBM Q team.


%

\end{document}